\documentclass[11pt]{article} 
\usepackage{graphicx,floatflt,amssymb,epsf,rotate} 
\begin{document}
\begin{center}
{\Large \bf
Perturbative generation of $\theta_{13}$ from tribimaximal neutrino mixing 
\\ }
\vskip 1cm
\renewcommand{\thefootnote}{\fnsymbol{footnote}}
Biswajoy Brahmachari$^1$\footnote{email: biswa.brahmac@gmail.com}
and Amitava Raychaudhuri$^2$\footnote{email:palitprof@gmail.com}\\
\vskip 1cm
(1) Department of Physics, 
Vidyasagar Evening College, \\
39 Sankar Ghosh Lane, Kolkata 700006, India\\
\vskip .5cm
(2) Department of Physics, 
University of Calcutta,\\ 92 Acharya Prafulla Chandra  Road, 
Kolkata 700009, India
\end{center}
\vskip 2cm
\thispagestyle{empty}
\begin{center}
\underbar{\bf Abstract}
\end{center}

Solar and atmospheric neutrino oscillations are consistent with a
tribimaximal form of the mixing matrix $U$ of the lepton sector.
Exact tribimaximal mixing leads to $\theta_{13}=0$. Recent
results from the Daya Bay and RENO experiments have established a
non-zero value of $\theta_{13}$. Keeping the leading behaviour of
$U$ as tribimaximal we perform a model-independent perturbative
calculation to incorporate a non-vanishing $\theta_{13}$.   We
identify the nature of the perturbation matrix and consider the
possibility of the solar neutrino splitting also resulting from
it. We calculate up to first order in perturbation theory and
evaluate the deviations proportional to $\sin \theta_{13}$ while
including CP-nonconservation.  Finally, we briefly discuss a
gauge model where such an addition to the neutrino mass matrix
arises through  one-loop effects.

\vskip 20pt

\noindent

\texttt{PACS No:~ 14.60.Pq} 

\texttt{Key Words:~~Neutrino, tribimaximal mixing }

\renewcommand{\thesection}{\Roman{section}} 
\setcounter{footnote}{0} 
\renewcommand{\thefootnote}{\arabic{footnote}} 
\noindent

\newpage

Experimental data of solar, atmospheric, accelerator, and
reactor neutrinos \cite{data} translate to information about
neutrino masses and mixing which can be summarised as
\cite{Gonzalez, Schwetz}:
\begin{eqnarray}
\Delta m_{21}^2 &=& (7.59 \pm 0.20) \times 10^{-5} \, {\rm eV}^2, \;\;
\theta_{12} = (34.4 \pm 1.0)^\circ, \nonumber \\
|\Delta m_{31}^2| &=& (2.46 \pm 0.12) \times 10^{-3} \, {\rm eV}^2, \;\; 
\theta_{23} = (42.8 ^{+4.7}_{-2.7})^\circ \nonumber \\
\theta_{13} &=& (5.6 ^{+3.0}_{-2.7})^\circ, \;\; \delta  \; {\rm unknown}.
\label{results}
\end{eqnarray}
These values of the mixing angles are consistent with 
a mixing matrix of tribimaximal form \cite{hps},
\begin{equation}
U^0=
\pmatrix{\sqrt{2 \over 3} & \sqrt{1 \over 3} & 0 \cr
-\sqrt{ 1 \over 6} & \sqrt{ 1 \over 3} & \sqrt{1 \over 2} \cr
\sqrt{ 1 \over 6} & -\sqrt{ 1 \over 3} & \sqrt{1 \over 2}},
\label{n1}
\end{equation}
which predicts the third mixing angle $\theta_{13}$ to be exactly
vanishing. 

Of late, the situation has taken a different turn. Results from
the Double Chooz \cite{DChooz} collaboration and more recently
the Daya Bay \cite{DayaBay} experiment indicate that
$\theta_{13}$ is, in fact, inconsistent with zero\footnote{Very
recently the RENO collaboration has measured $\sin^22\theta_{13}
= 0.113 \pm 0.013 \; {\rm (stat)} \pm 0.019\; {\rm (syst)}$
\cite{RENO}.}  by more than 5$\sigma$  - $\sin^22\theta_{13} =
0.092 \pm 0.016 \; {\rm (stat)} \pm 0.005\; {\rm (syst)}$
\cite{DayaBay}.  Therefore, in view of these significant findings, it
has to be concluded that the simple-minded tribimaximal picture
fails to adequately capture the observed neutrino mixing. The
smallness of $\theta_{13}$ compared to the other two mixing
angles encourages us to examine here whether the former could arise
from a small perturbation on the basic tribimaximal structure and
could lead to a realistic neutrino mixing matrix. 

We work in a flavor basis in which the charged lepton mass matrix
is diagonal\footnote{This fixes the singlet
right-handed charged leptons and the left-handed lepton doublets
in flavor space. In this basis, mixing in the lepton sector is determined
entirely by the neutrino mass matrix.}. If the left-handed
neutrino Majorana masses be $m_1,m_2,m_3$ then from eq.
(\ref{n1}) the mass matrix $M^0$, satisfying
tribimaximal mixing, when expressed  in the flavor basis has the
general form,
\begin{eqnarray}
M^0&=&{U^0}\pmatrix{m_1 && \cr &m_2&\cr && m_3}{U^0}^T = \pmatrix{
{2m_1+m_2 \over 3} & {m_2-m_1 \over 3} & {m_1 - m_2 \over 3}\cr
{m_2-m_1 \over 3} & {m_1+2m_2+3m_3 \over 6} & -{m_1+2m_2 - 3m_3 \over 6}\cr
{m_1-m_2\over 3} & -{m_1+2m_2-3m_3 \over 6} &
{m_1+2m_2 + 3m_3 \over 6}\cr
}\nonumber \\
&=& \pmatrix{
m_0 - {\Delta_{31} \over 3} & {(\Delta_{31} - \Delta_{32}) \over 3} &
-{(\Delta_{31} - \Delta_{32}) \over 3} \cr
{(\Delta_{31} - \Delta_{32}) \over 3} & m_0 + {\Delta_{31} \over 6} &
{(\Delta_{31} + 2\Delta_{32}) \over 6} \cr
-{(\Delta_{31} - \Delta_{32}) \over 3} & 
{(\Delta_{31} + 2\Delta_{32}) \over 6} & m_0 + {\Delta_{31} \over 6}  \cr
},
\label{unpert0}
\end{eqnarray}
where we have set
\begin{equation}
m_0 = (m_1 + m_2 + m_3)/3, \; \Delta_{32} \equiv (m_3 - m_2) \;
{\rm and} \;  \Delta_{31} \equiv (m_3 - m_1). 
\label{delta1}
\end{equation}
{\em Ab initio,} the mass eigenvalues, $m_1, m_2, m_3$, can be
complex in which case they can be rendered real and positive by a diagonal
phase transformation, $D = diag (e^{i\lambda_1}, e^{i\lambda_2},
1)$, where the $\lambda_i$ are Majorana phases, which do not
affect neutrino oscillations.

We approximate $\Delta_{32}  \simeq \Delta_{31} \equiv
\Delta$, which is not unreasonable since $|\Delta_{32}| \gg
\Delta_{21} \equiv (m_2 - m_1)$. $\Delta$ sets the scale for
atmospheric neutrino oscillations\footnote{$\Delta$ is positive
(negative) for the normal (inverted) ordering of neutrino
masses.}. We start with this limit and write the unperturbed mass
matrix  in the flavor basis as:
\begin{equation}
M^0 \simeq \pmatrix{
m_0 - {\Delta \over 3} & 0 & 0 \cr
0 & m_0 + {\Delta \over 6} & {\Delta \over 2}\cr
0 & {\Delta \over 2} & m_0 + {\Delta \over 6}  \cr}.
\label{unpert}
\end{equation}
At this level, $m^{(0)}_1 = m^{(0)}_2 = m_0 - {\Delta \over 3}$
and $m^{(0)}_3 = m_0 + {2\Delta \over 3}$ and the
solar mass splitting is absent.  Our goal is to also generate
this splitting through the same perturbation hamiltonian that is
responsible for $\theta_{13} \neq 0$. We take $m^{(0)}_1, m^{(0)}_2$,
and $m^{(0)}_3$ to be real and positive.

The purpose of this paper is not to explain how $M^0$ emerges 
from a fundamental model; even though there is no doubt that we 
consider it as the dominant part of the neutrino mass matrix. 
There are many models from which one can obtain the tribimaximal
form of the mixing matrix \cite{models}. Our discussion below will
be independent of the specific mechanism by which $M^0$ arises.

In terms of the three mixing angles  and the complex phase
$\delta$ the Pontecorvo, Maki, Nakagawa, Sakata (PMNS) mixing 
matrix is conventionally parametrized as, 
\begin{equation}
U=
\pmatrix{c_{12}c_{13} & s_{12}c_{13} & s_{13}e^{-i \delta} \cr
-s_{12}c_{23}-c_{12}s_{23}s_{13}e^{i \delta} & 
c_{12}c_{23}-s_{12}s_{23}s_{13}e^{i \delta} & 
s_{23}c_{13}
\cr
s_{12}s_{23}-c_{12}c_{23}s_{13}e^{i \delta} & 
-c_{12}s_{23}-s_{12}c_{23}s_{13}e^{i \delta} & 
c_{23}c_{13}
}.
\label{PMNS}
\end{equation}

As noted, the tribimaximal mixing matrix, $U^0$ in eq.
(\ref{n1}), fixes the element $U^0_{e3}=0$.  The r$\hat{\rm o}$le
of a non-vanishing $U_{e3}$, or equivalently $\theta_{13}$, is
manifold.  It is essential for CP-nonconservation  in neutrino
oscillations and may be invoked to explain
leptogenesis\footnote{The Majorana phases alluded to earlier
could produce CP-violation in $\Delta L = 2$ processes.}. Also,
$\theta_{13} \neq 0$ will  be similar to the quark sector where
mixing between all three generations and CP-violation is a
well-verified result, though the mixing angles in the two sectors
are vastly different. For CP-violation, of course, both
$\theta_{13}$ and the complex phase $\delta$ should be
non-vanishing. Besides, a reasonably large $\theta_{13}$ opens
the door for an easier measurement of the neutrino mass ordering,
i.e., the sign of $\Delta m_{31}^2$.

A large number of attempts have been made to generate
$\theta_{13} \neq 0$ in diverse ways starting from an initial
tribimaximal form. Some of these are the following. A
perturbative analysis in which one of the columns or rows of
$U^0$ is left unchanged has been examined in \cite{HeZee}. An
alternative which
involves  a sequential `integrating out' of heavy neutrino states
has been proposed in \cite{Grinstein}.  Another approach has been
to parametrize the deviation from the tribimaximal form in a
particular way \cite{unitary}. Deviations from tribimaximal
mixing due to charged lepton effects and Renormalization Group
running have been other directions of study \cite{Boudjemaa}.
Alternative explorations have been  based on the $A(4)$ symmetry in
\cite{A4pert, Adhikary}, and on other discrete symmetries in \cite{Delta,
Group}.

Our strategy here is to  use perturbation theory to identify the
structure of the  Majorana mass matrix, $M = M^0 + M'$, where $M'
\ll M^0$, so that  $\theta_{13}$  and the solar mass splitting
are obtained. Both $M^0$ and $M'$ will be symmetric and could, in
general, be complex. However, $M^0$ as obtained in eq.
(\ref{unpert}) from the tribimaximal mixing form is real and
symmetric, i.e., hermitian. We will consider the cases of real
and complex $M'$ separately.

For our later discussions the  eigenstates of $M^0$, the
unperturbed mass eigenstates, in the {\em mass} basis are found
useful. These are simply:
\begin{equation}
\psi^{(0)}_1 = \pmatrix{1 \cr 0 \cr 0}, ~~~
\psi^{(0)}_2 = \pmatrix{0 \cr 1 \cr
0}, ~~~ \psi^{(0)}_3 = \pmatrix{0 \cr 0 \cr 1},
\label{basis0}
\end{equation}
of which the first two are
degenerate. So, the basis vectors $\psi^{(0)}_1$ and
$\psi^{(0)}_2$ are not unique and are chosen with the knowledge
that they reproduce the correct solar mixing.  The physical
basis is fixed by the perturbation.  When we discuss
lifting of the degeneracy, we consider $M'$ to be such that
$\psi^{(0)}_1$ and $\psi^{(0)}_2$ are its non-degenerate
eigenstates: $<\psi^{(0)}_i|M^\prime|\psi^{(0)}_j> = m_i^{(1)}
\delta_{i j} ~(i,j = 1,2)$, with $m_1^{(1)} \neq m_2^{(1)}$. We
also take $(M^\prime)_{33} = 0$ in this mass basis, so what
remain are $(M^\prime)_{13}$ and $(M^\prime)_{23}$ to which we
will first turn.

It is helpful to bear in mind that eigenstates in eq.
(\ref{basis0}) when expressed in the {\em flavor} basis are simply
the columns of  $U^0$, eq. (\ref{n1}), namely,
\begin{eqnarray}
\psi^{(0)}_1 = \pmatrix{\sqrt{2 \over 3} \cr 
-\sqrt{1 \over 6} \cr \sqrt{1 \over 6} },~~~ 
\psi^{(0)}_2 = \pmatrix{\sqrt{1 \over 3} \cr 
\sqrt{1 \over 3} \cr -\sqrt{ 1 \over 3} }, ~~~
\psi^{(0)}_3 = \pmatrix{0 \cr \sqrt{1 \over 2} \cr \sqrt{1 \over
2} } \;\; {\rm in ~flavor ~basis}.
\label{basis}
\end{eqnarray}

The goal we have set ourselves is to obtain as the perturbed mass
eigenstates, when written in the flavour basis, the columns of
the matrix in eq. (\ref{PMNS}) with $\theta_{13} \neq 0$. To
this end, initially, let us take $M'$, which is symmetric,  to be
real and therefore hermitian.  Needless to say, this may generate
a non-zero $\theta_{13}$ but will have no CP-violation and hence
yield\footnote{Note that a negative $s_{13}$ with $\delta = 0$ is
equivalent to a positive $s_{13}$ and $\delta =
\pi$.} $\delta = 0$. For the perturbation expansion we retain
terms up to linear in $s_{13}$.  To first order  we have,
\begin{equation}
\psi_3=\psi^{(0)}_3+\sum_{j \neq 3} O_{3j}\psi^{(0)}_j \;.
\label{pert1}
\end{equation}
Here
\begin{equation}
O_{3j}={<\psi^{(0)}_j|M^\prime|\psi^{(0)}_3> \over
m^{(0)}_3-m^{(0)}_j}=-O_{j3},
\;\;\; (j \neq 3) .
\label{pert2}
\end{equation}
The coefficients $O_{3j}$ are real in this case. In the mass
basis $O_{ij}$ is proportional to $M_{ij}$. 

The eigenstate,
$\psi_3$, should correspond to the third column of the mixing
matrix $U$ in eq.  (\ref{PMNS}) with $\delta = 0$.  
$O_{31}$ and $O_{32}$ are readily determined using eq.
(\ref{pert1}) in the flavor basis.  Written explicitly  we get
the matrix equation,
\begin{equation}
\pmatrix{s_{13} \cr \cr s_{23}c_{13} \cr \cr c_{23}c_{13}}
= \pmatrix{0 & \sqrt{2 \over 3} & \sqrt{1 \over 3} \cr 
\sqrt{1 \over 2} & -\sqrt{1 \over 6} & \sqrt{1 \over 3} \cr
\sqrt{1 \over 2} & \sqrt{1 \over 6} & -\sqrt{1 \over 3} } 
\pmatrix{1 \cr \cr O_{31} \cr \cr O_{32}} \; .
\label{e1}
\end{equation}
By inverting the above equation one obtains, to order linear in
$s_{13}$, $O_{31} = \sqrt{2 \over 3}s_{13}$ and  $O_{32} =
\sqrt{1 \over 3}s_{13}$, where maximality of the atmospheric
mixing angle ($s_{23} = c_{23} = 1/\sqrt{2}$) has been used. This
translates to $M^\prime_{13} = \sqrt{2 \over 3}s_{13}\Delta $
and $M^\prime_{23} = \sqrt{1 \over 3}s_{13}\Delta $  in the mass basis.

To extend this discussion to the case of $\delta \neq 0$, we have
to bear in mind that now $M'$ is complex symmetric and {\em not}
hermitian and the same holds for the total Majorana mass matrix
$M = M^0 + M'$.  The columns of the mixing matrix $U$ (eq.
(\ref{PMNS})) are eigenvectors of
$M^\dagger M = M^{0\dagger}M^0 + M^{0\dagger}M' + M'^{\dagger}M^0$,
where we have dropped a term which is ${\cal O} (M')^2$. 
To proceed, we recall that $M^0$ is hermitian and
therefore the eigenstates of the unperturbed $M^{0\dagger}M^0$
are the same $\psi^{(0)}_i$ considered earlier -- eq.
(\ref{basis}) --  but now corresponding to eigenvalues
$(m^{(0)}_1)^2, (m^{(0)}_2)^2$ and $(m^{(0)}_3)^2$. In place of
eq.  (\ref{pert2}) we have
\begin{equation}
O_{3j}={<\psi^{(0)}_j| ( M^{0\dagger}M' + M'^{\dagger}M^0) |\psi^{(0)}_3> 
\over \left(m^{(0)}_3\right)^2 - \left(m^{(0)}_j\right)^2} =
-O^*_{j3}, \;\; (j \neq 3) 
\; ,
\label{pert3}
\end{equation}
which is to be used in eq. (\ref{pert1}) now. Requiring that
$\psi_3$ be reproduced to first order  and
using the appropriate variant of  eq.  (\ref{e1})  we get in
this case $O_{31} = \sqrt{2 \over 3}s_{13}e^{-i \delta}$ and  $O_{32} =
\sqrt{1 \over 3}s_{13}e^{-i \delta}$. 

To relate the above to the elements of the perturbation
$M^\prime$ one notes:
\begin{equation}
<\psi^{(0)}_j| ( M^{0\dagger}M' + M'^{\dagger}M^0) |\psi^{(0)}_i> 
 = m^{(0)}_j  <\psi^{(0)}_j| M'  |\psi^{(0)}_i>  + \;
m^{(0)}_i <\psi^{(0)}_j| M'^\dagger  |\psi^{(0)}_i> 
\; ,
\label{matelem}
\end{equation}
and thus  in the mass basis
\begin{equation}
O_{3j} \left[\left(m^{(0)}_3\right)^2 - \left(m^{(0)}_j\right)^2
\right] = m^{(0)}_j (M^\prime)_{j3} + m^{(0)}_3 (M^\prime)^*_{j3}
,\;\;   (j \neq 3) \; ,
\label{pert4}
\end{equation}
where the symmetric nature of $M^\prime$ has been used. Writing
$M^\prime_{13} = \sqrt{2 \over 3}s'_{13}\Delta ~e^{i \phi}$ and
$M^\prime_{23} = \sqrt{1 \over 3}s'_{13}\Delta ~e^{i \phi} $
using eq. (\ref{pert4}) one finds to leading order in
$\Delta/m_0$
\begin{equation}
\delta = \tan^{-1}\left( \frac {\Delta}{2 m_0} \tan\phi\right),
\;\; {\rm and} \;\; s_{13} = f(\phi) s'_{13}, 
\label{delta}
\end{equation} 
where 
\begin{equation}
f(\phi) = \frac{\left[ (m^{(0)}_1)^2 + (m^{(0)}_3)^2 + 2
m^{(0)}_1 m^{(0)}_3 \cos 2\phi \right]^{1/2}}{(m^{(0)}_1 +
m^{(0)}_3)} \;\;.
\end{equation}
The approximate formulae in (\ref{delta}) indicate that
$s_{13} \leq s'_{13}$, with the equality holding only when $\phi
= 0$, and  though the range of $\phi$ --  which is  $\{0,
2\pi\}$ -- is the same as that of $\delta$ the latter is
suppressed compared to the corresponding $\phi$. The suppression
is higher as the neutrino masses approach the quasi-degenerate
regime ($\Delta \ll m_0$).

So far we have concentrated on obtaining $\theta_{13} \neq 0$
 through a perturbation starting from the tribimaximal form. Now
we consider the solar mass splitting. We choose the perturbation
such that $(M^\prime)_{12} = (M^\prime)_{21} = 0$.  
The first order corrections to the neutrino mass are obtained from
$m^{(1)}_i= ~<\psi^{(0)}_i|M^\prime|\psi^{(0)}_i>$.
We demand that the following mass corrections arise at this order
\begin{equation}
m^{(1)}_1 = m^{(1)}_3 = 0 ~{\rm and} ~m^{(1)}_2 \neq 0.
\label{firstm}
\end{equation}
In the mass basis this implies that out of the diagonal elements
only $(M^\prime)_{22} \neq 0$. Such a correction  ensures that a
nonzero solar mass splitting $m_2 - m_1 = m^{(1)}_2$ is induced.
Solar neutrino observations establish $\Delta m^2_{21} = (m_2)^2
- (m_1)^2$ is positive.

Putting all this together we have for the full perturbation
matrix in the {\em mass} basis as
\begin{equation}
M^\prime = s'_{13} \Delta  \pmatrix{0 & 0 &\sqrt{2 \over 3}
~e^{i \phi} \cr
0 & x/3 & \sqrt{1 \over 3} ~e^{i \phi} \cr
\sqrt{2 \over 3} ~e^{i \phi} &
\sqrt{1 \over 3} ~e^{i \phi}
& 0}  \; {\rm ~in ~mass ~basis},
\label{pertallmb}
\end{equation}
and 
\begin{equation}
x \equiv m^{(1)}_2/s'_{13} \Delta\;.
\label{xdef}
\end{equation}
The
dimensionless parameter $x$ is fixed by the solar splitting.
In general it can be complex implying that the Majorana mass
$m^{(1)}_2 \equiv  |m^{(1)}_2| ~\exp(i \chi)$. If we write $m_2 =
m^{(0)}_2 + m^{(1)}_2 \equiv  |m_2| ~\exp(i \lambda)$ and recall
$m^{(0)}_1 = m^{(0)}_2$ then   one has:
\begin{equation} 
|m_2| = \left[(m^{(0)}_1)^2 + (|m^{(1)}_2|)^2 + 2 m^{(0)}_1 |m^{(1)}_2|
\cos \chi\right]^{1/2}, \;\;
\lambda = \tan^{-1}\left[{|m^{(1)}_2|  ~\sin\chi  \over
m^{(0)}_1 + |m^{(1)}_2|  ~\cos \chi}\right].
\label{m2}
\end{equation}   
$\lambda$ is a Majorana phase of $\nu_2$ which arises from the
perturbation. 

There are thus two real parameters introduced here: $|m^{(1)}_2|$
and $\chi$.  For any phase angle $\chi$ demanding that the solar
splitting is correctly obtained determines $|m^{(1)}_2|$ provided
$m^{(0)}_1$ is known, i.e., the mass ordering is specified and
the mass of the lightest neutrino, $\tilde m$, is given.  Thus
$\chi$, $s'_{13}$ and $\phi$ suffice to fix  the full
perturbation matrix $M'$.

Using $M^\prime$ in eq. (\ref{pertallmb}) and degenerate perturbation
theory \cite{Schiff} we
get for the mixing matrix with $\delta \neq 0$:
\begin{equation}
U_{\delta \neq 0}=
\pmatrix{\sqrt{2 \over 3} & \sqrt{1 \over 3} & {s_{13}e^{-i \delta}} \cr
-\sqrt{ 1 \over 6} - \sqrt{1 \over {3}}~s_{13} e^{i \delta} &
\sqrt{ 1 \over 3} - \sqrt{1\over {6}}~s_{13}e^{i \delta}  &
\sqrt{1 \over 2} \cr
\sqrt{ 1 \over 6} - \sqrt{1\over {3}}~s_{13}e^{i \delta}  &
-\sqrt{ 1 \over 3} - \sqrt{1\over {6}}~s_{13}e^{i \delta}  &
\sqrt{1 \over 2} } \;.
\label{upert2}
\end{equation}
$U_{\delta \neq 0}$ is consistent with the observed mixing angles
and is unitary up to order $s_{13}$. The non-zero CP-phase
$\delta$ brings the lepton sector in line with the quarks, where
CP-violation has been established for long. $\delta$ is usually invoked
for processes such as leptogenesis.  A matrix of exactly
the form of $U_{\delta \neq 0}$ has been discussed in
\cite{Xing} from a different motivation and its consistency with
the experimentally required mixing angles noted.

The basis independent measure of CP-violation, the leptonic
Jarlskog \cite{Jarlskog} invariant, arising from $U_{\delta \neq
0}$ (eq. (\ref{upert2})) is
\begin{equation}
J={\rm Im}[U_{e1}U_{\mu 2}U^*_{e2}U^*_{\mu 1}]=-{1 \over 3 \sqrt{2}}s_{13}
\sin \delta = -{1 \over 3 \sqrt{2}}s_{13}'  f(\phi) \frac{\left({\Delta
\over 2m_0}\right) \sin \phi}{\cos^2\phi + \left({\Delta
\over 2m_0}\right)^2 \sin^2 \phi } ,
\label{Jarls}
\end{equation}
signifying that both $s_{13}'$ and $\phi$ have to be non-vanishing
in order for CP-violation to be present in the lepton sector.
Moreover, in the quasi-degenerate regime the observation of
CP-violation is less likely.

The above discussion is valid when the solar mass splitting and
the mixing angle $\theta_{13}$ are unrelated.  In the following
we do not examine mass matrices of the associated general form --
eq.  (\ref{pertallmb}).   Nonetheless, we make one passing
remark. It would not be unreasonable to expect that the different
non-zero terms of the perturbation matrix (\ref{pertallmb}) are
rougly of similar order. We may then expect $x \sim {\cal O}(1)$.
Recalling eq.  (\ref{xdef}) one has the order of magnitude
estimate\footnote{This result is only indicative.  The full
flexibility of variation of $\phi$ and $\chi$ is not taken into
account.} $s_{13} \sim {\cal O}[(\Delta m^2_{21}/\Delta
m^2_{31}) (m_3^{(0)} + m_1^{(0)})/ (m_2^{(0)} + m_1^{(0)})]$. The
measured values of $\Delta m_{21}^2$ and $|\Delta m^2_{31}|$ are
known. We illustrate two extreme limits:  normal ordering with
$m_3^{(0)} \gg m_1^{(0)}, m_2^{(0)}$ implies $s_{13}
\sim {\cal O}[10^{-2} (m_3^{(0)}/2m_1^{(0)})]$ while for the
inverted ordering with $m_3^{(0)} \ll m_1^{(0)}, m_2^{(0)}$ one
has $s_{13} \sim {\cal O}[10^{-2}]$.  This is the
general expectation if both $\theta_{13}$
and the solar mass splitting arise from the same perturbation
of the tribimaximal mass matrix.

We now identify a special limit when the perturbation mass matrix
is of a texture which can be realised from a simple model and
where $s_{13}$ gets related to $\Delta m^2_{21}$ resulting in
restrictive predictions. To relate to mass models it is more convenient
to first rewrite $M^\prime$ in the flavor basis. We find
from eq.  (\ref{pertallmb})
\begin{equation}
M^\prime= s'_{13} \Delta \left[ \pmatrix{0 & \sqrt{1 \over 2} ~e^{i \phi}
& \sqrt{1 \over 2} ~e^{i \phi} \cr
\sqrt{1 \over 2} ~e^{i \phi} & 0 & 0 \cr
\sqrt{1 \over 2} ~e^{i \phi}
& 0 & 0} + \frac{1}{3} \pmatrix{x & x     &  - x  \cr
x  & x  & - x  \cr - x  & - x  & x }\right] \; \; {\rm ~in ~flavor ~basis}.
\label{pertall}
\end{equation}
Here the first matrix on the right-hand-side is responsible for
$\theta_{13}$ and the second for $\Delta m^2_{21}$.

We see from eq. (\ref{pertall})  that, aside from the diagonal
part which is proportional to the identity matrix\footnote{Such
a piece proportional to the identity does not affect the mixing
and makes a constant contribution to all three neutrino mass
eigenvalues.} and can be subsumed in $M^0$, the perturbation
is of the form:
\begin{equation}
M^\prime=\pmatrix{0 & A & B \cr A & 0 & C \cr B & C & 0}
\; \; {\rm ~in ~flavor ~basis},
\label{msimp}
\end{equation}
where $A$, $B$, and $C$ are complex in general. Such a texture of
$M'$  can  follow from a  Zee-type model \cite{zee} as we discuss
later. In such models $(M')_{\alpha \beta}$ is proportional to
$(m_\alpha^2 - m_\beta^2)$, where $m_\alpha$ is the mass of the
charged lepton $\alpha$. As $m_\tau \gg m_\mu , m_e$, unless other
couplings are of vastly different order from each other, one must have $B
\sim C \gg A$.   Such a form of
the mass matrix can be reproduced by the choice ${3 \over \sqrt
2}e^{i \phi} + x = \epsilon$, where $\epsilon$ is small, when  the
perturbation matrix eq. (\ref{pertall}) reduces to:
\begin{equation}
M^\prime=\frac{s'_{13}\Delta}{3 \sqrt 2} \pmatrix{0 & \epsilon
     & 6 e^{i \phi} + \epsilon  \cr
\epsilon & 0  & 3 e^{i \phi} - \epsilon  \cr
6 e^{i \phi} + \epsilon  & 3 e^{i \phi} - \epsilon  & 0 } \; .
\label{pertall2}
\end{equation}
The special case $\epsilon = 0$ is quite predictive.  From eq.
(\ref{xdef}) this requires
\begin{equation}
- {3 \over \sqrt 2} s'_{13} \Delta ~e^{i \phi} = m^{(1)}_2  =  
|m^{(1)}_2| ~e^{i \chi}.
\end{equation}
Thus $s'_{13} = \sqrt{2} ~|m^{(1)}_2| /3 |\Delta|$ and  $\phi =
\pi + \chi$ ($\phi = \chi$) for the normal (inverted) ordering.
   
Due to these relationships,  $\chi$ and $m^{(0)}_1$ besides
determining $|m^{(1)}_2|$ now also fix $s_{13}$ and $\delta$
through eq.  (\ref{delta}). As noted, $m^{(0)}_1$ is known when
the mass ordering and the lightest neutrino mass, $\tilde m$, are
fixed. So, for any mass ordering the two remaining parameters are
$\chi$ and the lightest neutrino mass. 

We show now that $\chi$ and $\tilde m$ can be chosen such that
one has consistency with both the solar mass splitting and the
measured $\theta_{13}$. In our discussion below we use the
central values of the atmospheric and solar mass splittings from
eq.  (\ref{results}) and seek an acceptable  $\theta_{13}$.  We
do not attempt an exhaustive listing of the entire consistent
ranges of the parameters in this work but rather present some
typical solutions for both mass orderings.

We find that with $\epsilon = 0$, in the normal mass ordering
case ($m^{(0)}_1 = \tilde m$) taking  $m^{(0)}_1 = 10^{-2}$ eV
and $115^\circ \leq \chi \leq 137^\circ$,  $\sin^22\theta_{13}$
varies from 0.057 to 0.130, which includes the 1$\sigma$
experimentally allowed range, while the Jarlskog CP-violation
parameter $J$ remains more or less constant around -0.026. The
replacement $\chi \leftrightarrow (2\pi - \chi)$ with $m^{(0)}_1$
fixed keeps $\sin^22\theta_{13}$ unchanged and replaces $J$ by
$-J$.

For the inverted mass ordering\footnote{In this case, $\tilde m =
m^{(0)}_3$ and $\Delta < $ 0.} ($m^{(0)}_1 = \tilde m - \Delta$),
on the other hand, taking $0 \leq m^{(0)}_3 \leq
10^{-3}$ eV and $\chi \sim 94^\circ$ one obtains
$\sin^22\theta_{13} \sim 0.054$, which is allowed at 2$\sigma$,
with $J \sim -0.028$. For $m^{(0)}_3 = 3 \times 10^{-2}$ eV and $\chi \sim
101^\circ$ one has $\sin^22\theta_{13} \sim 0.09$ within $1
\sigma$ and $J \sim
-0.029$. For any $m^{(0)}_3$, replacing  $\chi$ by $(2\pi -
\chi)$ results in the same  $\sin^22\theta_{13}$ but $J$ changes
sign.

\begin{figure} 
\begin{center} 
\epsfxsize=4in {\epsfbox{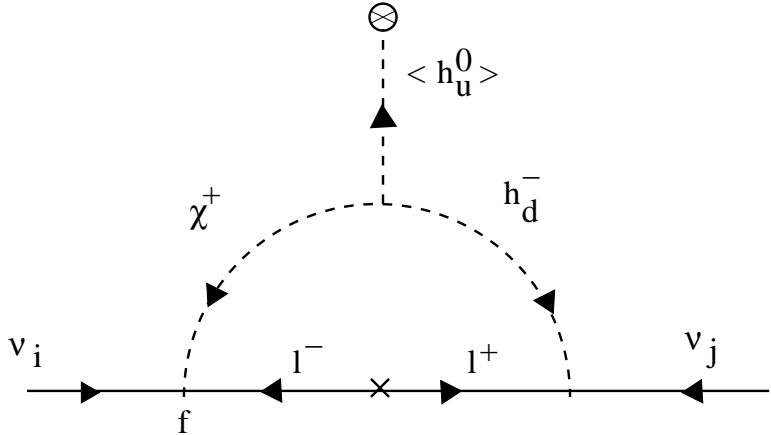}} 
\caption{\small One loop contributions to the neutrino Majorana
mass matrix $M'$ in the Zee model. For $M'_{e\mu}$ the dominant
contribution is proportional to $m_\mu$ while for $M'_{e\tau}$
and $M'_{\mu \tau}$  it is proportional to $m_\tau$.}
\label{o12} 
\end{center} 
\end{figure}

We now briefly note how $M'$ of the texture in eq. (\ref{msimp})
can   follow from a  Zee-type model\footnote{An
alternative way to generate a non-zero $\theta_{13}$ using the Zee
model has been examined in \cite{HeMajee}.}. It bears
repetition that here the Zee model provides a subleading
contribution, $M'$, to a leading tribimaximal mass matrix $M^0$
of a different origin\footnote{Models can be constructed which
accommodate both $M^0$ and $M'$. An $A(4)$ based example can be
found in \cite{Adhikary}.}.   The Zee model has a simple $SU(2)_L
\times U(1)_Y$ invariant structure.
For this,  a second scalar $SU(2)_L$ doublet and a charged
singlet scalar $\chi^+$ are introduced. The latter couples to a pair of
lepton doublets, where the coupling $f_{\alpha\beta}$ is
antisymmetric in the generation index. Likewise due to $SU(2)$
antisymmetry the charged scalar also couples to a pair of Higgs
doublets $h_u$ and $h_d$ antisymmetrically. In this model a
contribution to the neutrino mass -- $M'$ -- arises radiatively
from  one loop diagrams such as Fig. \ref{o12} and can
be expressed as:
\begin{equation}
M^\prime_{\alpha \beta}={1 \over
M_s^2}\mu (m^2_\alpha-m^2_\beta)f_{\alpha \beta}{v_u \over v_d} I\;.
\label{model}
\end{equation}
Here $f_{\alpha \beta}$ is the antisymmetric coupling in
$f_{\alpha \beta} L_\alpha L_\beta \chi$, where $L$ is the left
handed lepton doublet. Also, $\mu$ is the trilinear scalar
coupling in $\mu h_u h_d \chi$, $M_s$ a typical scalar mass, and
$I$ a dimensionless factor arising from the loop integral.  $v_u,
v_d$ are the vacuum expectation values of the two Higgs doublets
$h_u$ and $h_d$. The vertex $f$ violates lepton number by two
units.  This diagram gives rise to a mass matrix $M^\prime$ which
is off-diagonal and symmetric, as required.  $(M')_{12}$ can
be neglected compared to $(M')_{13}$ and $(M')_{23}$  because the
latter will receive contribution from diagrams with a
$\tau$-lepton\footnote{This indicates that the
general form of the perturbation matrix -- eq. (\ref{pertall2}) --
with arbitrary $\epsilon$ will require unnatural choices for the
couplings $f_{\alpha \beta}$.}. Thus, the correction obtained in
this fashion is naturally of the desired form with $\epsilon \sim
0$. Further, the
coupling $f_{\alpha \beta}$ can be complex which can lead to an
$M'$ of the form in eq. (\ref{msimp}) -- which is complex
symmetric. The interference of $M'$ with the matrix $M^0$ -- eq.
(\ref{pert3}) -- leads to  CP-violation in the neutrino
sector. It is worth bearing in mind that $M'$ is suppressed
compared to the leading term in $M^0$ by ${\cal O}(s_{13} \Delta /m)$.
Taking $s_{13} \sim 0.1$, $\Delta \sim 0.1$ eV and $\mu \sim 100$
GeV, unless other factors in eq. (\ref{model}) are tuned to suppress
the contribution, one requires $M_s \sim$ {$\cal O$}($10^6$ GeV),
which puts the additional scalars of the model beyond the reach
of the current experiments.

In conclusion, we have shown that $\theta_{13}$ consistent with
experiments, a CP-phase $\delta$, and the solar mass splitting
can all be the outcome of a specific perturbation to  a basic
neutrino mass matrix, the latter  associated with tribimaximal
mixing. This leads to a non-zero Jarlskog invariant and opens the
door for CP-violation in the lepton sector.  In particular, a
constrained version of this perturbation relates the neutrino
Majorana phase to the solar mass splitting as well as
$\theta_{13}$ and $\delta$. Some sample solutions which meet all
requirements have been presented.  We have provided an example
where the requisite perturbation contributions to the neutrino
Majorana mass matrix can arise from a Zee-type model through
radiative corrections.

\vskip 20pt
\noindent
{\bf Acknowledgements:} The research of AR is supported by a J.C.
Bose Fellowship of the Department of Science and Technology of
the Government of India.\\


\end{document}